\documentclass[final]{article}
\usepackage{geometry}  
\usepackage{amsmath,amssymb,amsfonts,graphicx}
\usepackage{algorithm}
\usepackage{cite}
\usepackage{geometry}
\usepackage{listings}
\usepackage{hyperref}
\usepackage{subcaption}
\usepackage{tikz}
\usepackage{caption}
\usepackage{algpseudocode}
\usepackage{hyperref}
\DeclareMathOperator\erf{erf}

\title{A Stochastic Performance Model for Pipelined Krylov Methods}
\author{Hannah Morgan \footnotemark[1] 
  \and Matthew G. Knepley \footnotemark[2]
  \and Patrick Sanan \footnotemark[3]
  \and L. Ridgway Scott \footnotemark[4]
}

\date{}

\begin{document}
\maketitle

\pagestyle{myheadings}
\thispagestyle{plain}
\markboth{}{MORGAN ET AL.} 

\renewcommand{\thefootnote}{\fnsymbol{footnote}}

\footnotetext[1]{Department of Computer Science, 1100 S. 58th St., University of Chicago, Chicago, IL 60637
  (hmmorgan@uchicago.edu), Supported by NSF grant OCI-1147680.}

\footnotetext[2]{Department of Computational and Applied Mathematics, Rice University, Houston TX 77005, Partial support from NSF grant OCI-1147680.}

\footnotetext[3]{Institute of Computational Science, Universit\`a della Svizzera italiana, 6904 Lugano, Switzerland.}

\footnotetext[4]{The Computation Institute and Departments of Computer Science and Mathematics, University of Chicago, Chicago, IL 60637, Partial support from NSF grants DMS-0920960 and DMS-1226019.}

\renewcommand{\thefootnote}{\arabic{footnote}}

\begin{abstract}
Pipelined Krylov methods seek to ameliorate the latency due to inner products necessary for projection by overlapping
it with the computation associated with sparse matrix-vector multiplication. We clarify a folk theorem that this can
only result in a speedup of $2\times$ over the naive implementation. Examining many repeated runs, we show that stochastic
noise also contributes to the latency, and we model this using an analytical probability distribution. Our analysis shows
that speedups greater than $2\times$ are possible with these algorithms.

\end{abstract}

%\begin{keywords}

\section{Introduction}

Krylov methods~\cite{saad2003} have become an indispensible tool for the scalable solution of large, sparse linear
systems, which in turn have enabled an explosion in massively parallel scientific
simulation~\cite{ScalesReport}. However, as we move to larger massively parallel architectures, the latency
cost for reduction operations, central to Krylov methods, has ballooned~\cite{HPCChallenge}. In an effort to control
these costs, ``pipelined'' versions of many Krylov algorithms have been developed, which allow some of the latency cost
to be hidden by computational work.

A pipelined version of the classical CG method was already developed in~\cite{Chronopoulos_Gear_1989} for vector
machines, revisited for large scale parallelism in~\cite{GhyselsVanroose2014}, and implemented for field-programmable
gate arrays in~\cite{StrzodkaGoddeke06}. Similarly, pipelined versions of CG~\cite{deSturler1995},
GMRES~\cite{GhyselsAshbyMeerbergenVanroose2013} and BiCGStab~\cite{JacquesNicolasVollaire12} have also been put forward.
One cannot arbitrarily remove synchronizations from algorithms and expect comparable behavior, so pipelined variants of Krylov methods employ rearrangements which give arithmetically equivalent methods with looser data dependencies and the possibility to overlap computation and global communication, at the cost of additional intermediate storage and local computation, increased latency as a pipeline is filled, and degraded numerical stability. We have included a GMRES algorithm (Algorithm~\ref{gmres_alg}) and a pipelined version by~\cite{GhyselsAshbyMeerbergenVanroose2013} (Algorithm~\ref{pgmres_alg}) below for completeness.

% side-by-side algorithms here
\noindent\begin{minipage}[t]{.5\textwidth}
\begin{algorithm}[H]
\caption{GMRES}\label{gmres_alg}
\begin{algorithmic}[1]
   \State $r_0 \gets b-Ax_0$; $v_0 \gets r_0/\|r_0\|_2$
\For{$i = 0, 1, \dots, m - 1$} 
\State{$z \gets Av_i$}
\State{$h_{j, i} \gets \langle z, v_j \rangle, j = 0, 1, \dots, i$}
\State{$\tilde{v}_{i + 1} \gets z - \sum_{j = 1}^i h_{j, i} v_j$}
\State{$h_{i+1, i} \gets \| \tilde{v}_{i + 1} \|_2$}
\State{$v_{i + 1}  \gets \tilde{v}_{i + 1}/h_{i+1, i}$}
\State{\# apply Givens rotations to $H_{:, i}$}
 \EndFor
\State $y_m \gets \text{argmin}\|(H_{m+1, m} y_m - \|r_0\|_2 e_1 )\|_2$
\State $x \gets x_0 + V_m y_m$
\vspace{1.637in}
\end{algorithmic}
\end{algorithm}
\end{minipage}%
\begin{minipage}[t]{.5\textwidth}
\begin{algorithm}[H]
 \caption{PGMRES}\label{pgmres_alg}
\begin{algorithmic}[1]
   \State $r_0 \gets b-Ax_0$; $v_0 \gets r_0/\|r_0\|_2$; $z_0 \gets v_0$
\For{$i = 0, 1, \dots, m + 1$} 
\State $w \gets Az_i$
\If{i $>$ 1}
\State $v_{i-1} \gets v_{i-1}/h_{i-1, i-2}$
\State $z_i \gets z_i/h_{i-1, i-2}$
\State $w \gets w/h_{i-1, i-2}$
\For{$j = 0, 1, \dots, i$}
\State{$h_{j, i-1} \gets h_{j, i-1}/h_{i-1, i-2}$}
\EndFor
\State $h_{i-1, i-1} \gets h_{i-1, i-1}/h^2_{i-1, i-2}$
\EndIf
\State{$z_{i + 1} \gets w - \sum_{j = 0}^{i-1} h_{j, i-1} z_{j+1}$}
\If{i $>$ 0}
\State{$v_i \gets z_i - \sum_{j = 0}^{i-1} h_{j, i-1} v_j$}
\State $h_{i, i-1} \gets \|v_i\|_2$
\EndIf
\State{$h_{j, i} \gets \langle z_{i+1}, v_j \rangle, j = 0, 1, \dots, i$}
\EndFor
\State $y_m \gets \text{argmin}\|(H_{m+1, m} y_m - \|r_0\|_2 e_1 )\|_2$
\State $x \gets x_0 + V_m y_m$
\end{algorithmic}
\end{algorithm}
\end{minipage}

It has been difficult to understand the performance of these solvers on existing machines, judge the impact of
algorithmic tradeoffs, and predict performance on future architectures due to the lack of a coherent performance
model~\cite{HoeflerLumsdaineRehm07}. For example, some runs in~\cite{GhyselsVanroose2014} exhibit a speedup of slightly more than a factor of 2, but this
is difficult to explain in a deterministic model, as will be shown in Section~\ref{sec:model}.

In this work, we present a stochastic performance model for pipelined Krylov solvers, detailed in
Section~\ref{sec:model}, and examine the implications of different waiting time distributions on algorithm performance
in Section~\ref{sec:stochastic}. These predictions are compared to parallel experiments in Section~\ref{sec:exp}.

\section{Mathematical Model}\label{sec:model}

We model a Krylov iterative method as a set of $P$ communicating processes who must perform a calculation consisting of
local computations, separated by periodic global synchronizations, and interrupted by waiting, perhaps due to
unsatisfied requests to memory, actions of the operating system, etc. We will label the set of computations and
waiting by the index $k$, which corresponds to the iteration number for the Krylov method. The removal of the global
synchronizations will correspond to the introduction of split-phase collectives~\cite{HoeflerLumsdaineRehm07} for the
norm calculation and orthogonalization step. A split-phase, or non-blocking, collective is a collective operation, such
as a broadcast, which has been split into two parts so that it no longer requires a global synchronization at the point
of the call. Rather, the collective operation is first initiated, say with \texttt{MPI\_Ibcast()}, and then later
finalized with \texttt{MPI\_Wait()}. This strategy removes the global synchronization and allows the operation to be
overlapped with useful work between the initialization and finalization. The ratio between times with and without
synchronization will give us a bound on the speedup of the pipelined algorithm over the classical variant. Below we
illustrate this model using $P=2$ processes.

\subsection{Deterministic Computation and Waiting Times}

\begin{figure}
\tikzstyle{comp} = [rounded corners,fill=green]
\tikzstyle{wait} = [rounded corners,fill=purple]
\begin{tikzpicture}[scale=0.75]
  \path (0, 0) node[anchor=east] {$p=0$} (0, 1) node[anchor=east] {$p=1$};
  \foreach \x in {0, 6.3, 12.6}
  {
    \draw[comp] (\x, -0.25) rectangle +(3, 0.5) ++(0, 1) rectangle +(3, 0.5);
    \draw[wait] (\x, -0.25) ++(3.1, 0) rectangle +(3, 0.5) ++(0, 1) rectangle +(1.5, 0.5);
    \draw[ultra thick,dashed] (\x, -0.5) ++(6.2, 0) -- +(0, 2);
  }
\end{tikzpicture}
\caption{Computation with $P = 2$ processes for $K = 3$ steps. The green rectangles represent computation, the
  purple waiting, and the dotted lines are synchronization.\label{fig:comp}}
\end{figure}
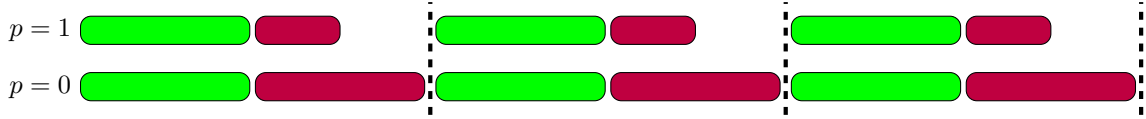

In the simplest scenario, each process takes a certain time $c_p$ for local computation, $w_p$ for waiting, which is
independent of the step $k$. Fig.~\ref{fig:comp} represents computation with three steps.
The total running time $T$, or \textit{makespan}, of the computation is then given by the expression
\begin{align}
  T = \sum_k \max_p (c_p + w_p) = K\, \max_p T_p,
\end{align}
where $K$ is the total number of steps, and $T_p = c_p + w_p$ is the time on process $p$ for one step excluding waiting for the
global barrier. Clearly, without loss of generality, we can replace the separate computation and waiting timing with a
single process time $T_p$.

\begin{figure}
\tikzstyle{comp} = [rounded corners,fill=green]
\tikzstyle{wait} = [rounded corners,fill=purple]
\begin{tikzpicture}[scale=0.75]
  \path (0, 0) node[anchor=east] {$p=0$} (0, 1) node[anchor=east] {$p=1$};
  \foreach \x in {0, 6.3, 12.6}
  {
    \draw[comp] (\x, -0.25) rectangle +(3, 0.5);
    \draw[wait] (\x, -0.25) ++(3.1, 0) rectangle +(3, 0.5);
  }
  \foreach \x in {0, 4.8, 9.6}
  {
    \draw[comp] (\x, 0.75) rectangle +(3, 0.5);
    \draw[wait] (\x, 0.75) ++(3.1, 0) rectangle +(1.5, 0.5);
  }
\end{tikzpicture}
\caption{Computation with $P = 2$ processes for $K = 3$ steps. The green rectangles represent computation, the
  purple waiting, and there is no synchronization.\label{fig:nosync}}
\end{figure}
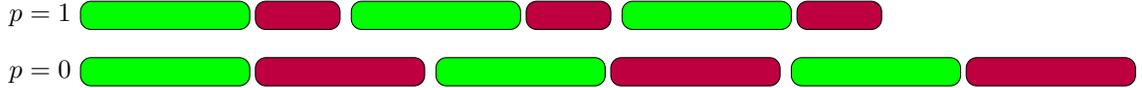

If we remove the synchronizations, as shown in Fig.~\ref{fig:nosync},
then the makespan is given by
\begin{align}
  T' = \max_p \sum_k (c_p + w_p) = K \max_p T_p,
\end{align}
so that no speedup is achievable. The removal of synchronizations can in general be modeled by the interchange of the
sum over steps and the maximum over process times.

\subsection{Stochastic Process Times}

If we allow the amount of local computation and waiting to fluctuate over the steps, as shown in Fig.~\ref{fig:variableWait}, we will see that speedup is achievable, as
shown in Fig.~\ref{fig:variableWaitNoSync}.
%\begin{figure}
%\tikzstyle{comp} = [rounded corners,fill=green]
%\begin{tikzpicture}[scale=0.75]
%  \path (0, 0) node[anchor=east] {$p=0$} (0, 1) node[anchor=east] {$p=1$};
%  \draw[comp] (0, -0.25) rectangle +(1, 0.5) ++(0, 1) rectangle +(5, 0.5);
%  \draw[ultra thick,dashed] (0, -0.5) ++(5.1, 0) -- +(0, 2);
%  \draw[comp] (5.2, -0.25) rectangle +(5, 0.5) ++(0, 1) rectangle +(1, 0.5);
%  \draw[ultra thick,dashed] (0, -0.5) ++(10.3, 0) -- +(0, 2);
%\end{tikzpicture}
%\caption{Computation with $P = 2$ processes for $K = 2$ steps. The green rectangles represent computation.\label{fig:variableComp}}
%\end{figure}
Most simulation codes, including most linear solvers, statically partition the data so that computation times do not show
large fluctuations. Waiting times, however, are more volatile and fluctuate across processes and
steps~\cite{SkinnerKramer05}, which can arise from interactions with the OS~\cite{FerreiraBridgesBrightwell08}.

In the very simple scenario that one process waits for a long time $W$ on the first step, the other on the second, and
on other steps the processes both take time $T_0$, as shown in Fig.~\ref{fig:variableWait},
\begin{figure}
\tikzstyle{comp} = [rounded corners,fill=green]
\tikzstyle{wait} = [rounded corners,fill=purple]
\begin{tikzpicture}[scale=0.75]
  \path (0, 0) node[anchor=east] {$p=0$} (0, 1) node[anchor=east] {$p=1$};
  \draw[comp] (0, -0.25) rectangle +(1, 0.5) ++(0, 1) rectangle +(1, 0.5);
  \draw[wait] (0,  0.75) ++(1.1, 0) rectangle +(5, 0.5);
  \draw[ultra thick,dashed] (0, -0.5) ++(6.2, 0) -- +(0, 2);
  \draw[comp] (6.3, -0.25) rectangle +(1, 0.5) ++(0, 1) rectangle +(1, 0.5);
  \draw[wait] (6.3, -0.25) ++(1.1, 0) rectangle +(5, 0.5);
  \draw[ultra thick,dashed] (6.3, -0.5) ++(6.2, 0) -- +(0, 2);
  \foreach \x in {12.6, 13.8, 15}
  {
    \draw[comp] (\x, -0.25) rectangle +(1, 0.5) ++(0, 1) rectangle +(1, 0.5);
    \draw[ultra thick,dashed] (\x, -0.5) ++(1.1, 0) -- +(0, 2);
  }
\end{tikzpicture}
\caption{Computation with $P = 2$ processes for $K = 5$ steps. The green rectangles represent computation, the
  purple waiting, and the dotted lines are synchronization.\label{fig:variableWait}}
\end{figure}
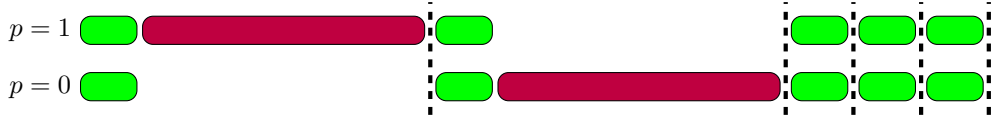
\begin{figure}
\tikzstyle{comp} = [rounded corners,fill=green]
\tikzstyle{wait} = [rounded corners,fill=purple]
\begin{tikzpicture}[scale=0.75]
  \path (0, 0) node[anchor=east] {$p=0$} (0, 1) node[anchor=east] {$p=1$};
  % p = 0
  \draw[wait] (0, -0.25) ++(2.2, 0) rectangle +(5, 0.5);
    \foreach \x in {0, 1.1, 7.3, 8.4, 9.5}
  {
    \draw[comp] (\x, -0.25) rectangle +(1, 0.5);
  }
  % p = 1
  \draw[wait] (0, 0.75) ++(1.1, 0) rectangle +(5, 0.5);
   \foreach \x in {0, 6.2, 7.3, 8.4, 9.5}
  {
    \draw[comp] (\x, 0.75) rectangle +(1, 0.5);
  }
\end{tikzpicture}
\caption{Computation with $P = 2$ processes for $K = 5$ steps. The green rectangles represent computation, the
  purple waiting, and there is no synchronization.\label{fig:variableWaitNoSync}}
\end{figure}
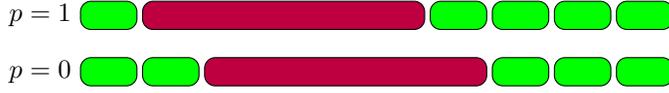
then the makespan with and without synchronizations is given by
\begin{align}
  T  &= \sum_k \max_p T_p = 2 W + K T_0,\\
  T' &= \max_p \sum_k T_p = W + K T_0.
\end{align}
Thus the possible speedup is
\begin{align}
  \frac{T}{T'} = \frac{2 W + K T_0}{W + K T_0} = \frac{2 + \alpha}{1 + \alpha}
\end{align}
where $\alpha = K T_0/W$, which is bounded above by 2. Extended to $P$ processes this gives an upper bound of $P$ on
the speedup, since all the waiting time is collapsed to the first interval when synchronization is removed and
computation time becomes small. This is related to the folklore result that speedup from covering communication with
computation is limited to a factor of 2, where we have two players, computing and communicating. This is because if we
cover all communication by computation, then one is larger than the other and is more than half of the computational time.

\section{Stochastic Model}\label{sec:stochastic}

We will employ a stochastic description of the waiting time variation making it amenable to analysis. We begin
with a stochastic process time $\mathcal{T}^k_p$ at each step that is drawn from a distribution independent of
process and stationary in step number. Let $T$ be the total Krylov time (computation and waiting) of the classical algorithm with synchronization, and $T'$ the
total time for the pipelined version without synchronization where $T = \sum_k \max_p \mathcal{T}^k_p$ and $T' = \max_p \sum_k \mathcal{T}^k_p$. We can ask for the expected total time with and without synchronizations,

\begin{align}
  E[T] = E[\sum_k \max_p \mathcal{T}^k_p] = \sum_k E[\max_p \mathcal{T}^k_p], \label{eq:sync_time}
\end{align}
and
\begin{align}
  E[T'] = E[\max_p \sum_k \mathcal{T}^k_p]. \label{eq:async_time}
\end{align}
These stage times will be modeled as random variables, drawn from some underlying distribution. Since the time spent
computing should not fluctuate, within our measurement accuracy, it only affects the mean of the distribution. The
variability in the distribution models the waiting times which can arise from interactions with the
OS~\cite{SkinnerKramer05,FerreiraBridgesBrightwell08}. We assume that waiting times across processes are independent
because we have no expectation that OS operations will be correlated across processes. In this section, we compute the
ratio of these quantities for a range of representative distributions for waiting times, in order to estimate the
potential speedup.

\subsection{Formulation}

We would like to calculate the speedup after $k$ steps, $\frac{E[T]}{E[T']}$,
where expected total time $E[T]$ is defined by~\eqref{eq:sync_time} and likewise $E[T']$ is defined
by~\eqref{eq:async_time}. First, we will derive an expression for the expected value of the maximum of a set of random variables in order to find an expression for $E[T]$. Then we will find one for $E[T']$.

A general expression for the expected value of the maximum of a set of random variables can be found in~\cite{Pitman93}
and is described here for completeness.  Let $X_1, X_2, \dots, X_n$ be independent, identically distributed (iid) random
variables with probability distribution function (pdf) $f(x)$ and cumulative distribution function (cdf) $F(x)$.  Let $X_{\max}$ be
another random variable such that $X_{\max} = \max\{X_1, X_2, \dots, X_n\}$.  By noticing that $X_{\max} \leq x$ if and
only if $X_i \leq x$ for all $i$'s,
\begin{align} \nonumber
F_{\max}(x) &= P(X_{\max} \leq x) \\ \nonumber
&= P(X_1 \leq x, X_2 \leq x, \dots, X_n \leq x) \\ \nonumber
&= P(X_1 \leq x) P( X_2 \leq x) \dots P( X_n \leq x) \\ \nonumber
&= F(x) F(x) \dots F(x) \\ \nonumber
&= F(x)^n \\ \nonumber
\end{align}
using the fact that the $X_i$'s are independent and identically distributed.  Furthermore, 
\begin{align} 
f_{\max}(x)= \frac{d}{dx}F_{\max}(x) = \frac{d}{dx} F(x)^n = n F(x)^{n-1} f(x) \nonumber
\end{align}
since by definition $\frac{d}{dx} F(x) = f(x)$.  The expected value of $X_{\max}$, integrating over the support of $x$, is
\begin{align} 
E[X_{\max}] = n \int^{\infty}_{-\infty} x F(x)^{n-1} f(x) dx. \label{E-max}
\end{align}

% before Charlie's comments
% At step $k = 1$, we have $P$ iid random variables $\mathcal{T}^1_0, \mathcal{T}^1_1, \dots, \mathcal{T}^1_{P-1}$ from an underlying distribution with pdf $f_1(x)$, cdf $F_1(x)$, and joint cdf $F(x_0, x_1, \dots, x_{P-1})$. Similarly, for a given step $k$, $\mathcal{T}^k_0, \mathcal{T}^k_1, \dots, \mathcal{T}^k_{P-1}$ are iid random variables with pdf $f_k(x)$, cdf $F_k(x)$, and joint cdf $F(x_0, x_1, \dots, x_{P-1})$. Note that the joint cdfs are the same at both steps since the random variables $\mathcal{T}^k_p$ are stationary in $k$.  Thus, $\prod_i F_1(x_i) = \prod_i F_k(x_i)$ for all $x_i$. In particular, this is true when $x_i = x$ for all $i$, so that $F_1(x)^P = F_k(x)^P$, which gives us $F_1(x) = F_k(x)$. This also implies that $f_1(x) = f_k(x)$ so that, using~\eqref{E-max}, we have

% after Charlie's comments
At step $k = 1$, we have $P$ iid random variables $\mathcal{T}^1_0, \mathcal{T}^1_1, \dots, \mathcal{T}^1_{P-1}$ from an underlying distribution with pdf $f_1(x)$, cdf $F_1(x)$, and joint cdf $F(x_0, x_1, \dots, x_{P-1})$. Similarly, for a given step $k$, $\mathcal{T}^k_0, \mathcal{T}^k_1, \dots, \mathcal{T}^k_{P-1}$ are iid random variables with pdf $f_k(x)$, cdf $F_k(x)$, and joint cdf $F(x_0, x_1, \dots, x_{P-1})$. Since the random variables are stationary in $k$, the joint cdf and thus the individual pdfs and cdfs remain the same at step $k$ so that $f_1(x) = f_k(x)$ and $F_1(x) = F_k(x)$. Then using~\eqref{E-max}, we have

 \begin{align} 
E[\max_p \mathcal{T}^1_p] = P \int^{\infty}_{-\infty} x F_1(x)^{P-1} f_1(x) dx = E[\max_p \mathcal{T}^k_p]. \nonumber %\label{eq:T1-max}
\end{align}
Then
\begin{align} \nonumber
  E[T] = \sum_k E [\max_p \mathcal{T}^k_p] = K E [\max_p \mathcal{T}^1_p].  \nonumber
\end{align}

Also because the random variables $\mathcal{T}^k_p$ are stationary in $k$, $E[\mathcal{T}^k_p] = \mu$ for all $k$. The sum $\sum_k  \mathcal{T}^k_p$ will approach $K\mu$ in the limit of large $K$ so that we also have
\begin{align} \nonumber
  E[T'] &= E \big [\max_p  \sum_k  \mathcal{T}^k_p \big] \to K \mu. \nonumber
\end{align}
In this case, speedup is given by $\dfrac{E[T]}{E[T']} \to \dfrac{E [\max_p \mathcal{T}^1_p]}{\mu}$.

We will examine a of range of common analytical distributions. As the tails of $f_k(x)$ become heavier, the
potential speedup increases and can eventually exceed $2\times$.

\subsection{Uniform Distribution}
Let the $\mathcal{T}_p$'s from $P$ processes be independent random variables from a uniform distribution on $[a, b]$ with pdf $f(x) = \frac{1}{b-a}$, cdf $F(x) = \frac{x-a}{b-a}$, and mean $\mu = \frac{a+b}{2}$.  Using \eqref{E-max}, we  calculate the expected value of the maximum
\begin{align} \nonumber
E[\max_p \mathcal{T}_p] &= P \int^{b}_{a} x \bigg(\frac{x-a}{b-a}\bigg)^{P-1}\frac{1}{b-a} dx \\ \nonumber
&= \frac{a+Pb}{P+1}
\end{align}
to find speedup on $P$ processes
\begin{align}
\frac{E[T]}{E[T']}  =\dfrac{\dfrac{a+Pb}{P+1}}{\dfrac{a+b}{2}} =\frac{2(a+Pb)}{(P+1)(a+b)}. \nonumber
\end{align}
On the interval $[0, b]$, we find that the asynchronous speedup, $\dfrac{2P}{P+1}$, is bounded from above by 2.

\subsection{Exponential Distribution}
Let $\mathcal{T}_0, \mathcal{T}_1, \mathcal{T}_2, \mathcal{T}_3$, the times for four processes, be independent random variables from an exponential distribution with pdf $f(x) = \lambda e^{-\lambda x}$, cdf $F(x) = 1 - e^{-\lambda x}$, and mean $\mu = \frac{1}{\lambda}$.  Again, using \eqref{E-max}, we calculate the expected value of the maximum
\begin{align} \nonumber
E[\max_p \mathcal{T}_p] &= 4 \lambda  \int^{\infty}_{0} x \big(1 - e^{-\lambda x}\big)^{3} e^{-\lambda x} dx \\ \nonumber
&= \frac{25}{12 \lambda}.
\end{align}
We find that the speedup on four processes is
\begin{align}
\frac{E[T]}{E[T']}  =\dfrac{25/12 \lambda}{1/ \lambda} =\frac{25}{12} > 2. \nonumber
\end{align}
When the $\mathcal{T}_p$'s are from an exponential distribution, asynchronous speedup is greater than 2 on four or more processes.  Furthermore, the speedup on $P$ processes 
\begin{align}
\frac{E[T]}{E[T']}  =\dfrac{E[\max_p \mathcal{T}_p]}{\mu} =\frac{ \lambda P \int^{\infty}_{0} x \big(1 - e^{-\lambda x}\big)^{P-1} e^{-\lambda x} dx}{1/\lambda}  = H_P.\nonumber
\end{align}
Here, $H_P = \log P + \gamma + O(1/P)$ is the $P$th harmonic number and $\gamma$ is Euler's constant ~\cite{Scott2011}.

\subsection{Log-normal Distribution}
Let the $\mathcal{T}_p$'s from $P$ processes be independent random variables from a log-normal distribution with
pdf $f(x)=\frac{1}{x \sqrt{2 \pi} \sigma} e^{\frac{-(\ln(x) - \mu)^2}{2 \sigma^2}}$, cdf $F(x) =
\frac{1}{2}+\frac{1}{2}\erf\big(\frac{\ln(x)-\mu}{\sqrt2 \sigma}\big)$, and mean $\mu' = e^{\mu + \frac{\sigma^2}{2}}$.
Note that if a random variable $X$ is normally distributed with mean $\mu$ and variance $\sigma$, then $Y = e^X$ is a
random variable from a log-normal distribution with mean $\mu'$. After fixing $P$, $\mu$, and $\sigma$, we can calculate
the speedup from $P$ processes numerically using equation \eqref{E-max} and Octave's {\tt quad} function.  First, let
$P=2$, $\mu = 0$, and $\sigma = 1$.  Equation \eqref{E-max} becomes
\begin{align} \nonumber
  E[\max_p \mathcal{T}_p] = 2 \int^{\infty}_{0} x\bigg( \frac{1}{2}+\frac{1}{2}\erf\bigg(\frac{\ln(x)}{\sqrt2}\bigg)\bigg) \bigg(\frac{1}{x \sqrt{2 \pi}} e^{\frac{-(\ln(x))^2}{2}}\bigg) dx \approx 2.5069, \nonumber
\end{align}
so that the speedup on two processes is
\begin{align}
  \frac{E[T]}{E[T']}  =\dfrac{E[\max_p \mathcal{T}_p]}{\mu'} \approx \frac{2.5069}{\sqrt{e}} \approx 1.5205. \nonumber
\end{align}
Now let $P=4$, $\mu = 0$, and $\sigma = 1$.  Equation \eqref{E-max} becomes
\begin{align} \nonumber
  E[\max_p \mathcal{T}_p] = 4 \int^{\infty}_{0} x\bigg( \frac{1}{2}+\frac{1}{2}\erf\bigg(\frac{\ln(x)}{\sqrt2}\bigg)\bigg)^3 \bigg(\frac{1}{x \sqrt{2 \pi}} e^{\frac{-(\ln(x))^2}{2}}\bigg) dx
\approx 3.6406. \nonumber
\end{align}
We again find that on four processors, the potential speedup is greater than 2:
\begin{align}
  \frac{E[T]}{E[T']} \approx \frac{3.6406}{\sqrt{e}} \approx 2.2081 > 2. \nonumber
\end{align}

%\hannah{quad can have an infinite endpoint by using Gauss-Kronrod or tanh-sinh integration.}  

%\subsection{Comparison to Experiment}

%\begin{itemize}
%  \item Model for GMRES should have a growing mean
%\end{itemize}

\section{Experimental Results}\label{sec:exp}

% Section on Goodness-of-Fit tests: http://www.itl.nist.gov/div898/handbook/prc/section2/prc21.htm

The speedups reported in~\cite{GhyselsVanroose2014} for PIPECG and PIPECR on an Intel Xeon cluster using Infiniband seem
limited by $2\times$ speedup, but at the limit of 20 processes exceed this slightly (2.09 and 2.14
respectively). However, understanding the origin of speedup exceeding $2\times$ requires the examination of many
identical runs in order to amass statistics. Thus, we have generated repeated runs for PETSc KSP tutorial ex23 using CG,
PIPECG, GMRES, and PGMRES on 8192 processors of the Piz Daint Cray XC30 supercomputer at CSCS~\cite{PizDaint}, storing
the data in an open repository~\cite{Sanan15}. The PIPECG and PGMRES algorithms are similar in that work involving
sparse matrix-vector products (SpMV) is juxtaposed with work involving vector dot products. The dot products are reductions
which require some sort of global synchronization. Thus the computational portion of our model is associated to the SpMV
and orthogonalization with the BLAS AXPY calls, whereas the synchronizations apply to the dot product portion. These
algorithms decouple these two operations, so that as long as the SpMV is more expensive than a global synchronization,
the dot product operation involves negligible waiting.

The ex23 tutorial uses a simple, tridiagonal system of size 2,097,152, which is a one-dimensional discretization of the
Laplacian, and we force 5000 iterates of the Krylov method. The pipelined methods produce almost identical
residuals to the original methods for this problem. Most of the runtime for CG and PIPECG is thus concentrated
in dot products, the \texttt{VecTDot()} operation in PETSc, rather than in SpMV. This means there is no computation to
cover the communication cost, and very often we see no speedup, or even slowdown, for these
runs. In~\cite{GhyselsVanroose2014}, PIPECG achieves $2\times$ speedup for SNES tutorial ex48 because the much denser
matrices in ex48 provide enough computation to cover the communication costs. PETSc ex48 solves the hydrostatic, that is
Blatter-Pattyn, equations for ice sheet flow, where the ice uses a power-law rheology with Glen exponent 3. This
generates a much denser system of equations with about 10x more nonzeros per row than ex23. The GMRES and PGMRES runs
are also constrained to use 5000 iterates, however here this generates considerable work in the orthogonalization
phase. This is analogous to the situation with PIPECG for ex48, so it can be covered by the split collective, and we
again see a roughly $2\times$ speedup. We note that an increase in the number of iterates is possible, but not observed
in~\cite{GhyselsAshbyMeerbergenVanroose2013,GhyselsVanroose2014} for the setup we use in this paper.

%Analysis for Patrick Data (Hannah says that these look similar to the overall time)
%\begin{itemize}
%  \item Distribution of VecTDot times in PIPECG
%  \item Distribution of VecDot() and KSPGMRESOrthog() times in PGMRES.
%\end{itemize}

%TODO Note that exponential distribution is ``memoryless''

However, we also see some outliers in the data where speedup exceeds $2\times$. This could potentially be explained by
assuming a noise distribution and using the prior results from Sec.~\ref{sec:stochastic}. Thus, we would like to show
that a set of observed random variables $X_1, X_2, \dots, X_n$ comes from some underlying  distribution using well-known statistical tests. In our case, we want to fit two different sets of observations: the multiples of twelve (so that
$n=12$) runs of a pipelined GMRES algorithm and twenty runtimes ($n=20$) of a pipelined CG algorithm. Both algorithms
were run on 8192 processors on Piz Daint. 

\subsection{Cram\'er-von Mises}

Let $X_1, X_2, \dots, X_n$ be $n$ observed values in increasing order and assume that $F(x)$ is the cdf of the
underlying distribution. The Cram\'er-von Mises test statistic is given by
\begin{align}
  T = \frac{1}{12n} + \sum_{i = 1}^n \bigg[\frac{2i-1}{2n} - F(X_i)\bigg]^2
\end{align}
If $T$ is larger than a tabulated limit value, we reject our assumption that the observations came from the distribution
with cdf $F(x)$. \emph{P}-values for the test statistic can be found in~\cite{CsorgoFaraway1996} and critical values
in~\cite{RigdonBasu2000}. We use a significance level $\alpha = 0.05$ in all of our tests.

Using the Cram\'er-von Mises statistic, we look at the consistency of our observations with both uniform and exponential
distributions. The cdfs are given by $F(x) = \frac{x - a}{b - a}$ over the support of $x$ and $F(x) = 1-e^{(-\lambda
  x)}$ for $x>0$, respectively. The Cram\'er-von Mises test allows one to estimate the parameters of the distribution
from the sample, unlike non-parametric tests such as Kolmogorov-Smirnov where the underlying distribution parameters are assumed to
be known. In the case of the uniform distribution, we will let the parameters $a$ and $b$ be $X_1$ and $X_n$, the
minimum and maximum observations and we will use maximum-likelihood estimation (MLE) to recover $\lambda$ for the
exponential distribution, which in this case gives $\frac{1}{\lambda} = \frac{1}{n} \sum_{i = 1}^{n} X_i$.

\subsection{Lilliefors}

The Lilliefors test is a normality test based on Kolmogorov-Smirnov where the expected value and the variance of the
underlying distribution are not specified. We will use it to fit our observations to a log-normal distribution by first
taking the natural logarithm of each sample $X_1, X_2, \dots, X_n$ and normalizing each sample so that
\begin{align}
  Z_i = \frac{\ln(X_i) - (\bar{x})}{s}
\end{align}
where $\bar{x}$ is the sample mean and $s$ the sample standard deviation. The Lilliefors test statistic can then be
calculated using
\begin{align}
  T = \sup \big| F(x) - S(x) \big|
\end{align}
where $F(x)$ is the standard normal cdf and $S(x)$ is the empirical distribution function for $Z_1, Z_2, \dots,
Z_n$. Again, if the test statistic $T$ is larger than a tabulated critical value, we reject the hypothesis that the
sample came from a normal distribution with significance level $\alpha$. Critical values can be found
in~\cite{RigdonBasu2000} and we use the Matlab function \texttt{lillietest} to calculate $T$.

\subsection{PGMRES and PIPECG}

\begin{center}
\begin{figure}[h]
\begin{center}
 \begin{subfigure}[b]{0.3\textwidth}
     \hspace*{-.8cm} \includegraphics[width=\textwidth, angle=-90]{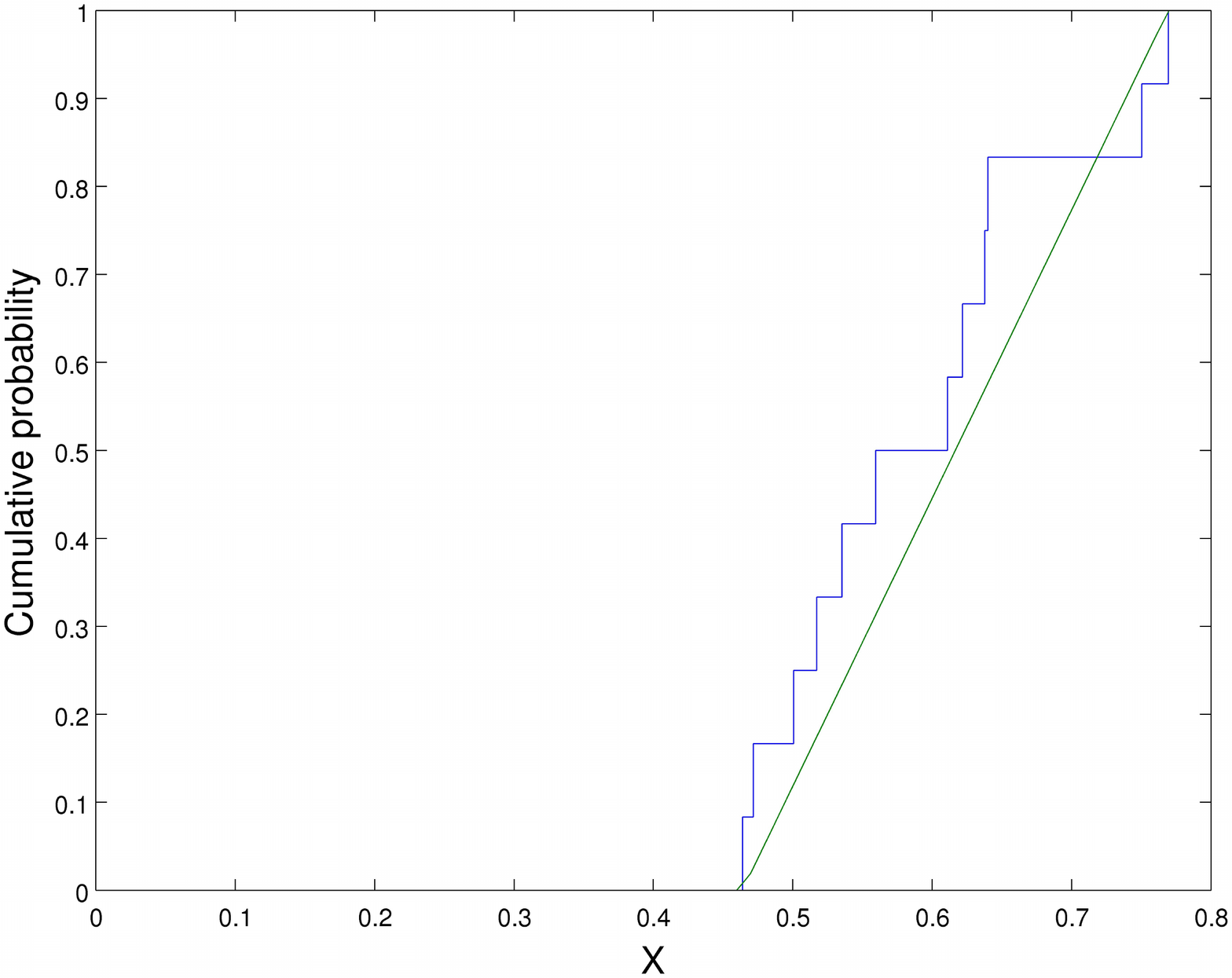}
     \caption{Uniform Distribution\label{fig:pgmres:a}}
  \end{subfigure} \quad \quad \quad
    \begin{subfigure}[b]{0.3\textwidth}
    \hspace*{-.8cm} \includegraphics[width=\textwidth, angle=-90]{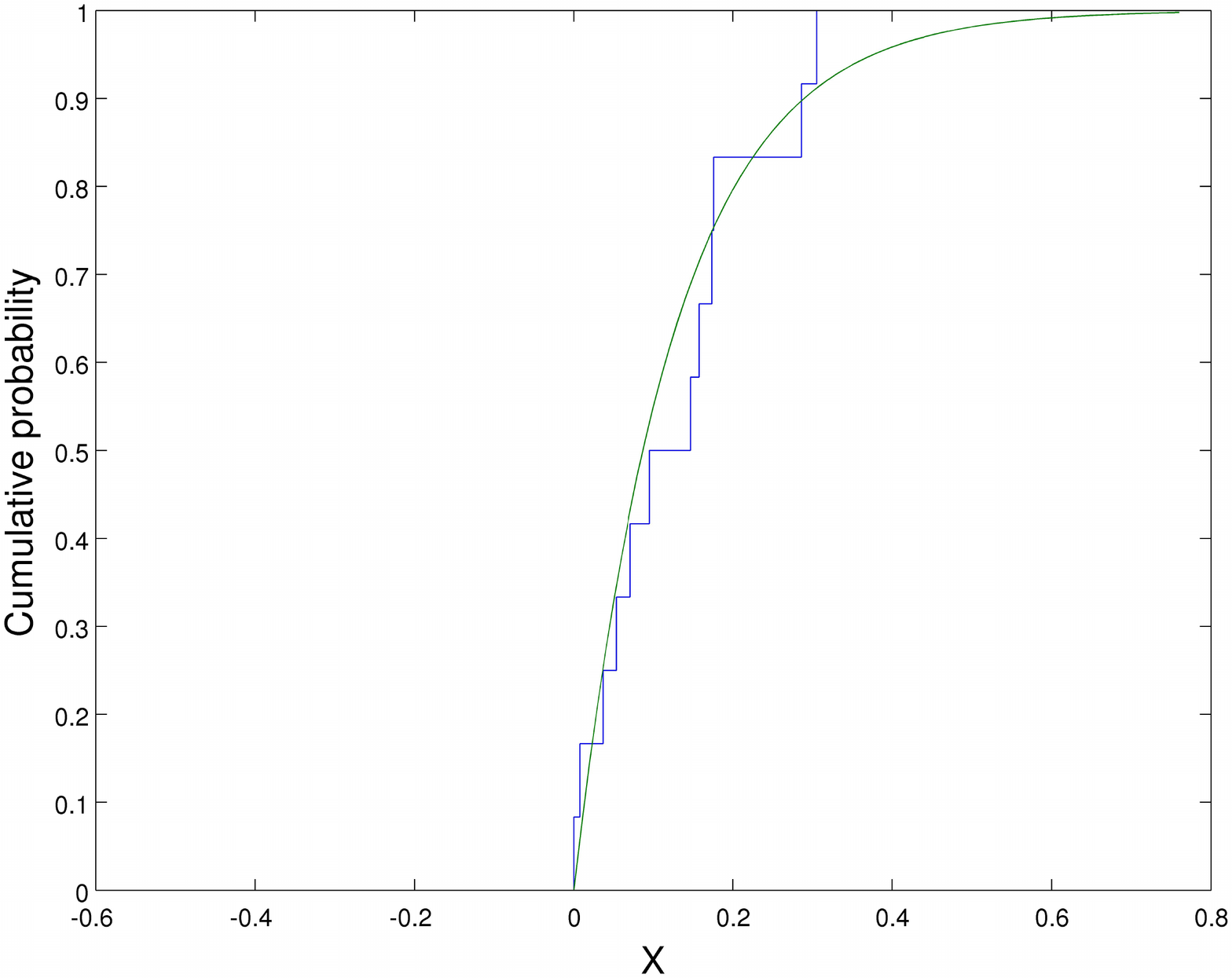}
      \caption{Exponential Distribution\label{fig:pgmres:b}}
  \end{subfigure} \quad \quad \quad
     \begin{subfigure}[b]{0.3\textwidth}
    \hspace*{-.8cm} \includegraphics[width=\textwidth, angle=-90]{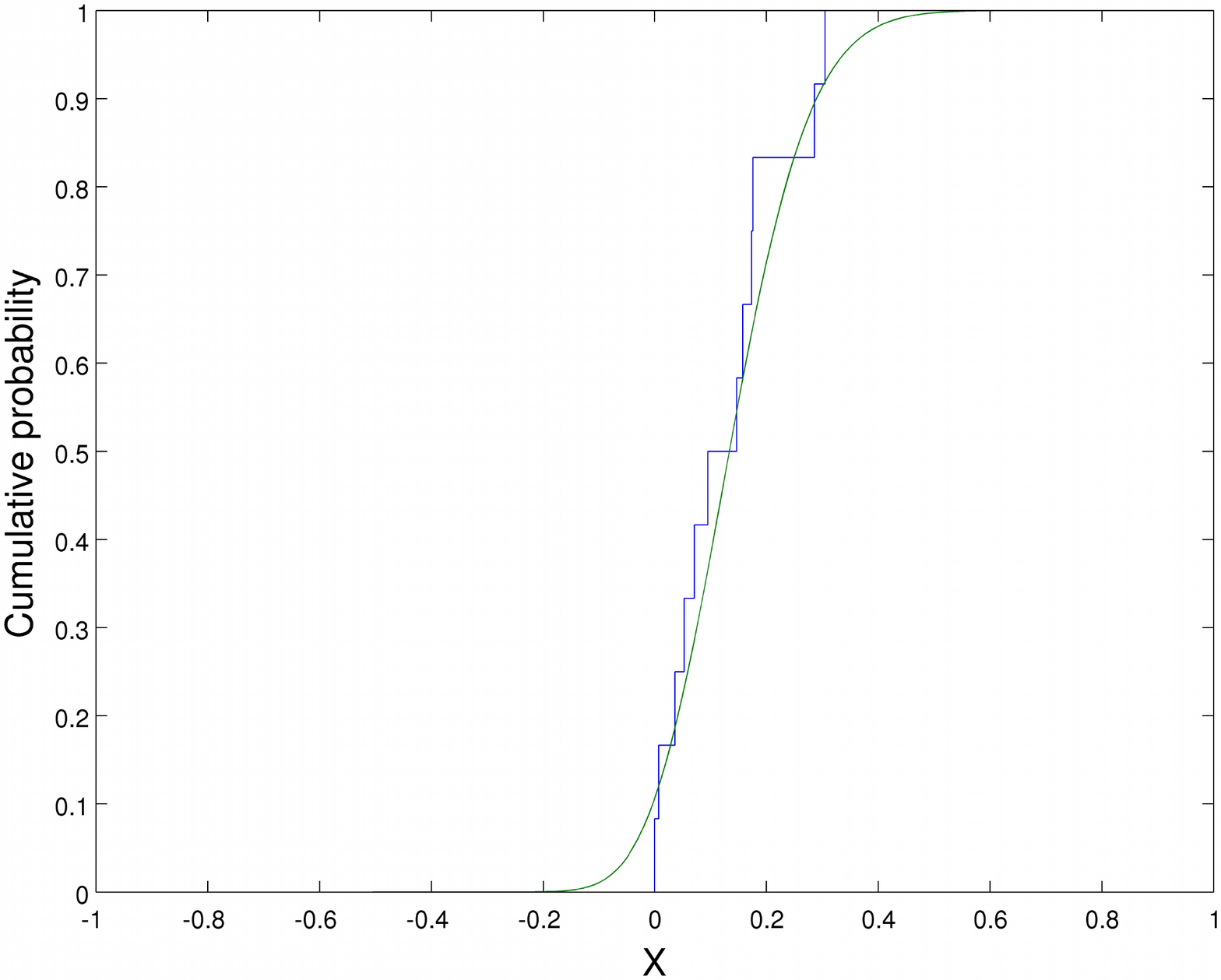}
    \caption{Log-normal Distribution\label{fig:pgmres:c}}
  \end{subfigure}
  \caption{We plot the empirical cumulative distribution for running times of PGMRES on PETSc KSP ex23, and also the MLE
    fit for analytic distributions,~\ref{fig:pgmres:a} uniform,~\ref{fig:pgmres:b} exponential, and~\ref{fig:pgmres:c}
    log-normal. \label{fig:pgmres}}
  \end{center}
\end{figure}
\end{center}

\begin{figure}[h!]
\centering
\begin{center}
 \begin{subfigure}[b]{0.3\textwidth}
 \hspace*{-.8cm} \includegraphics[width=\textwidth, angle=-90]{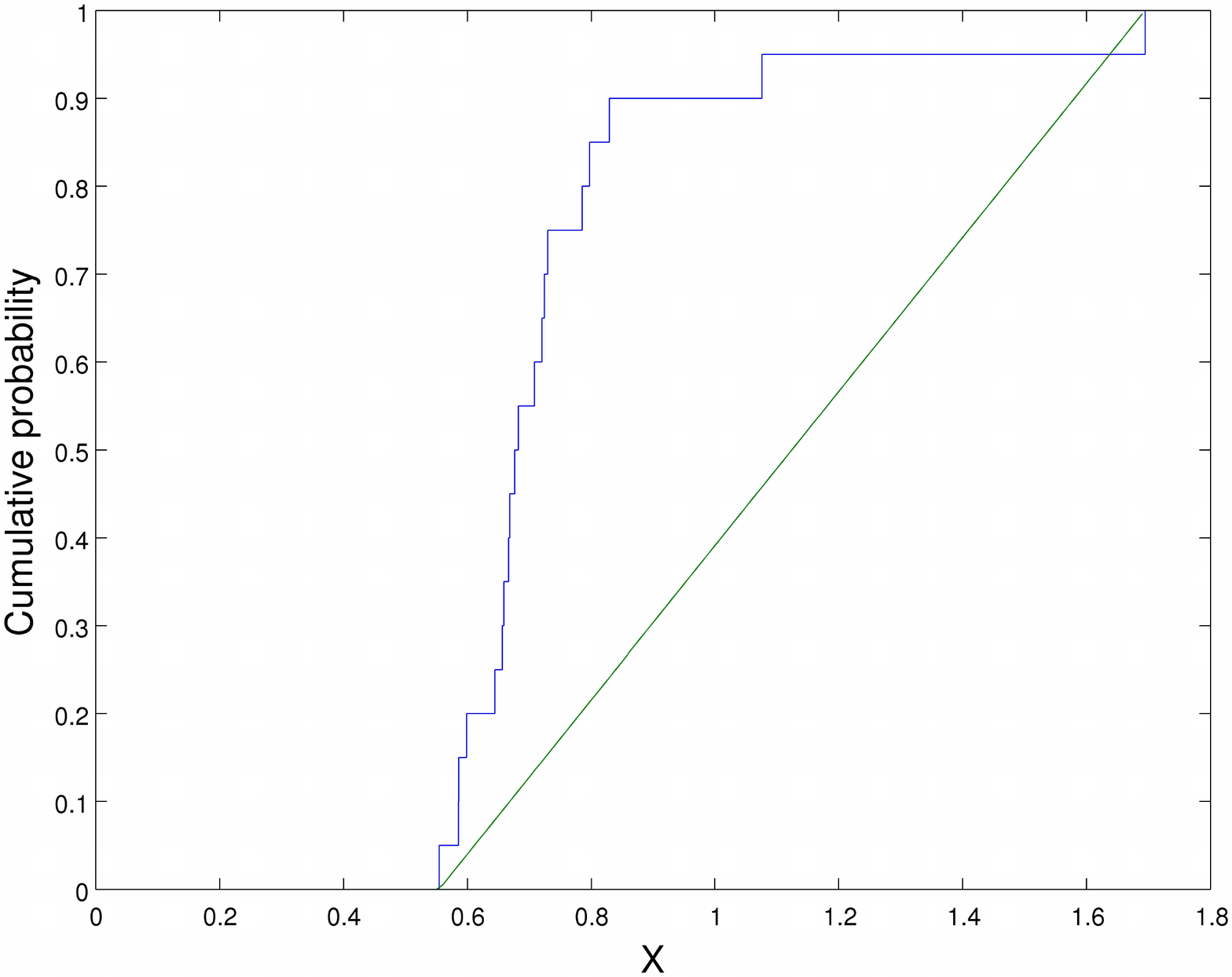}
    \caption{Uniform Distribution \label{fig:pipecg:a}}
  \end{subfigure} \quad \quad \quad
    \begin{subfigure}[b]{0.3\textwidth}
 \hspace*{-.8cm} \includegraphics[width=\textwidth, angle=-90]{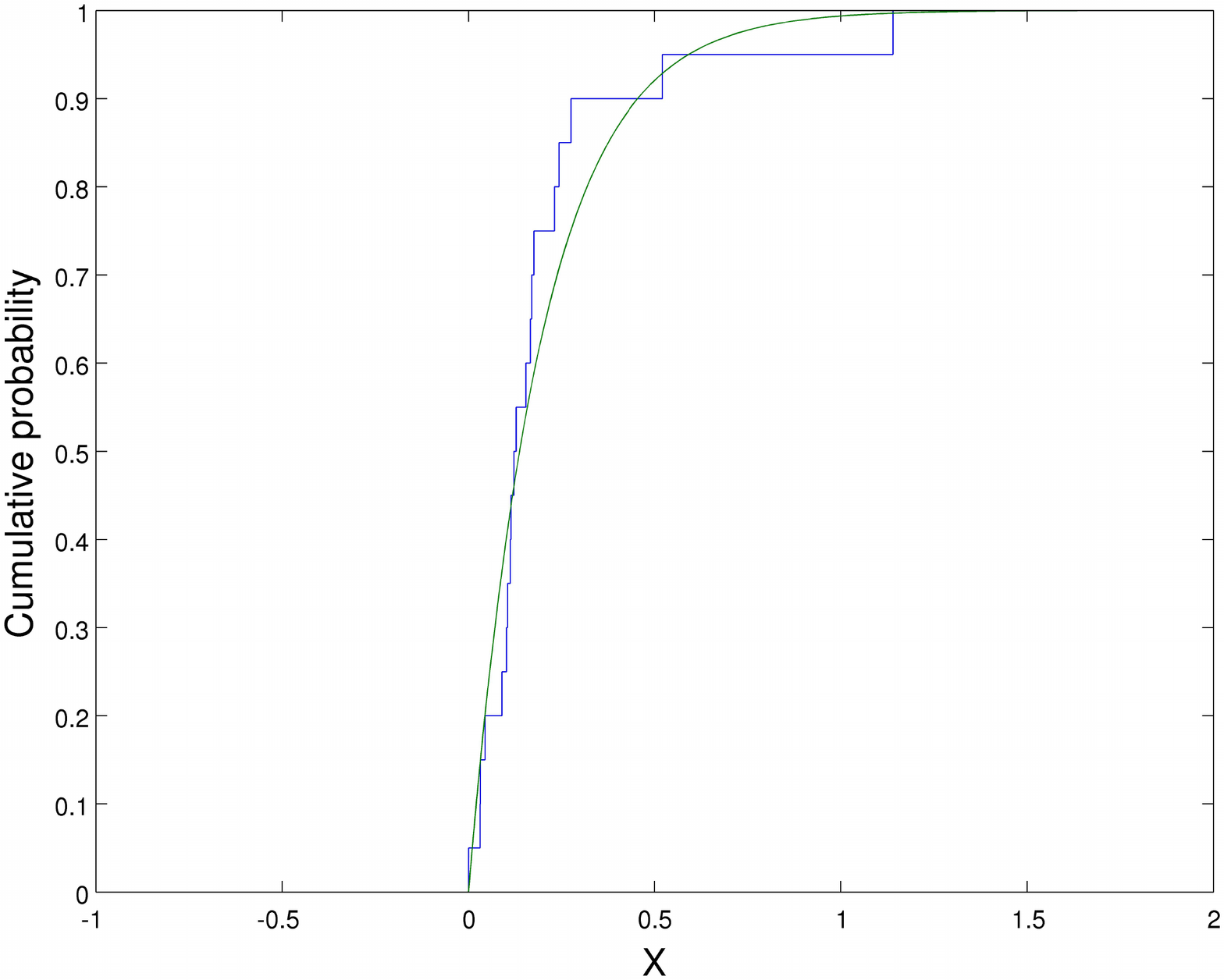}
    \caption{Exponential Distribution\label{fig:pipecg:b}}
  \end{subfigure}
  %\caption{Empirical PIPECG data with assumed underlying distributions. \label{fig:pipecg}}
  \caption{We plot the empirical cumulative distribution for running times of PIPECG on PETSc KSP ex23, and also the MLE
    fit for analytic distributions,~\ref{fig:pipecg:a} uniform, and~\ref{fig:pipecg:b} exponential. \label{fig:pipecg}}
  \end{center}
\end{figure}

Summary statistics for the GMRES, PGMRES, CG, and PIPECG runs are given in Table~\ref{tab:stats}, and the empirical cumulative
distribution functions along with MLE proposed distributions are shown in Fig.~\ref{fig:pgmres} and~\ref{fig:pipecg}.

\begin{table}
\caption{PGMRES and PIPECG runtime statistics \label{tab:stats}}
\centering
%\tabsize
\begin{tabular}{ l | c | c | c | c }
\hline
& GMRES & PGMRES & CG & PIPECG \\
\hline
$\bar{x}$      & 0.9465  & 0.5902   & 0.9349  & 0.7521 \\
median         & 0.9932  & 0.5856   & 0.8632  & 0.6792 \\
s              & 0.1303  & 0.0962   & 0.2385  & 0.2429 \\
$\text{s}^2$   & 0.0170  & 0.0092   & 0.0569 & 0.0590 \\
$\lambda$      & 1.0565  & 1.6942   & 1.0696  & 1.3295 \\
$X_{\text{min}}$ & 0.6617 & 0.4644   & 0.6051  & 0.5545 \\
$X_{\text{max}}$ & 1.0740 & 0.7697   & 1.6060  & 1.6950 \\
\hline
\end{tabular}  
\end{table}

Considering Fig.~\ref{fig:pgmres}, with significance level $\alpha = 0.05$, we reject the assumption that the PGMRES
runtimes come from a uniform distribution, and clearly the data are quite far from uniform. Using the Cram\'er-von Mises
statistic, we cannot reject the hypothesis that the distribution is exponential, nor can we reject a log-normal
distribution using the Lilliefors test. Due to the small variation in the runs we cannot distinguish these alternatives
with any confidence. More experiments must be run in order to obtain better statistics.

Using the same methodology in Fig.~\ref{fig:pipecg}, it is clear that the PIPECG runtimes are not uniform. Like PGMRES,
most of the samples are clustered in a small range, but one outlier ran for over twice as long as the sample
median. Using our tests, we reject that the runtimes come from uniform and log-normal distributions, but they are 
consistent with an exponential distribution.

It appears that some level of system noise is present in these simulations, since the exponential distribution provides
a much better explanation for the data than a narrow uniform window, and this noise enables significant speedup when
using the asynchronous Krylov methods. In Sec.~\ref{sec:stochastic} we showed that for exponentially distributed noise,
speedup is given by
\begin{align}
 H_P =  \sum_{i = 1}^{P} \frac{1}{p}
\end{align}
on $P$ processors. Thus PIPECG could possibly attain speedup greater than 2 when $P \geq 4$, but it appears in practice
that many more processors are necessary.

\section{Conclusions}\label{sec:conc}

According to the performance data in publication dealing with asynchronous Krylov methods~\cite{GhyselsVanroose2014}
that overall speedup did not exceed $2\times$, confirming the folk theorem for the case that we do no more than overlap
communication with computation. However, looking at a large number of repeated runs, we find that operating system noise
makes a measurable contribution, and can raise the speedup bound necessitating a new analysis. We show that uncommon
delays, which could quite possibly have been rejected as outliers in other work, are well-modeled by an exponential
distribution. Combined with our new analysis, this demonstrates that speedup greater than $2\times$ is possible for isolated
runs, and in a statistical sense.

In future work, we will apply these algorithms to high latency situations where we expect unpredictable delays, such as
heavily loaded machines, loosely coupled networks such as those employed for cloud computing or wireless computing
clusters, and heterogeneous machines such as those using GPUs. These workloads must be seen in a statistical sense since
most users will see computations contaminated by unpredictable, stochastic delays. Our analysis shows the effectiveness
of asynchronous methods in these situations, the necessity of characterizing the distribution of noise, and can perhaps
guide performance tuning of current algorithms and the development of new projection methods.

We also note that other potential sources of speedup have been identified. In~\cite{RuppWeinbubJungelGrasser15}, the
authors show that pipelined solvers can be implemented on a GPU using fewer kernel launches. Since kernel launch incurs
a significant latency penalty on the GPU, pipelined solvers can realize speedup from this fact alone.

\bibliographystyle{wileyj}
\bibliography{petsc,petscapp,document}

\begin{thebibliography}{10}
\providecommand{\url}[1]{\texttt{#1}}
\providecommand{\urlprefix}{URL }
\expandafter\ifx\csname urlstyle\endcsname\relax
  \providecommand{\doi}[1]{doi:\discretionary{}{}{}#1}\else
  \providecommand{\doi}{doi:\discretionary{}{}{}\begingroup
  \urlstyle{rm}\Url}\fi

\bibitem{saad2003}
Saad Y. \emph{Iterative Methods for Sparse Linear Systems, 2nd edition}. SIAM:
  Philadelpha, PA, 2003.

\bibitem{ScalesReport}
Keyes DE ( (ed.)). \emph{A Science-based Case for Large-scale Simulation}. U.S.
  Department of Energy, 2004. \urlprefix\url{http://www.pnl.gov/scales}.

\bibitem{HPCChallenge}
Dongarra J, Luszczek P, \emph{et~al.}. {HPC} challenge 2015.
  \urlprefix\url{http://icl.cs.utk.edu/hpcc/hpcc_results_lat_band.cgi?display=opt}.

\bibitem{Chronopoulos_Gear_1989}
Chronopoulos AT, Gear CW. {$s$-step} iterative methods for symmetric linear
  systems. \emph{Journal of Computational and Applied Mathematics}  1989;
  \textbf{25}:153--168.

\bibitem{GhyselsVanroose2014}
Ghysels P, Vanroose W. Hiding global synchronization latency in the
  preconditioned conjugate gradient algorithm. \emph{Parallel Computing}  2014;
  \textbf{40}(7):224--238, \doi{http://dx.doi.org/10.1016/j.parco.2013.06.001}.
  \urlprefix\url{http://www.sciencedirect.com/science/article/pii/S0167819113000719},
  7th Workshop on Parallel Matrix Algorithms and Applications.

\bibitem{StrzodkaGoddeke06}
Strzodka R, G\"oddeke D. Pipelined mixed precision algorithms on {FPGA}s for
  fast and accurate {PDE} solvers from low precision components.
  \emph{Proceedings of the 14th Annual IEEE Symposium on Field-Programmable
  Custom Computing Machines}, IEEE Computer Society, 2006; 259--270. FCCM '06.

\bibitem{deSturler1995}
de~Sturler E, van~der Vorst H. Reducing the effect of global communication in
  {GMRES}(m) and {CG} on parallel distributed memory computers. \emph{Applied
  Numerical Mathematics}  1995; \textbf{18}(4):441--459.

\bibitem{GhyselsAshbyMeerbergenVanroose2013}
Ghysels P, Ashby T, Meerbergen K, Vanroose W. Hiding global communication
  latency in the {GMRES} algorithm on massively parallel machines. \emph{SIAM
  Journal on Scientific Computing}  2013; \textbf{35}(1):C48--C71.

\bibitem{JacquesNicolasVollaire12}
Jacques T, Nicolas L, Vollaire C. Electromagnetic scattering with the boundary
  integral method on {MIMD} systems. \emph{High-Performance Computing and
  Networking}, \emph{Lecture Notes in Computer Science}, vol. 1593. Springer,
  1999; 1025--1031.

\bibitem{HoeflerLumsdaineRehm07}
Hoefler T, Lumsdaine A, Rehm W. Implementation and performance analysis of
  non-blocking collective operations for {MPI}. \emph{Proceedings of the 2007
  International Conference on High Performance Computing, Networking, Storage
  and Analysis, SC07}, IEEE Computer Society/ACM, 2007.

\bibitem{SkinnerKramer05}
Skinner D, Kramer W. Understanding the causes of performance variability in
  {HPC} workloads. \emph{Workload Characterization Symposium, 2005. Proceedings
  of the IEEE International}, IEEE, 2005; 137--149.

\bibitem{FerreiraBridgesBrightwell08}
Ferreira KB, Bridges P, Brightwell R. Characterizing application sensitivity to
  {OS} interference using kernel-level noise injection. \emph{Proceedings of
  the 2008 ACM/IEEE Conference on Supercomputing}, SC '08, IEEE Press:
  Piscataway, NJ, 2008; 19:1--19:12.
  \urlprefix\url{http://dl.acm.org/citation.cfm?id=1413370.1413390}.

\bibitem{Pitman93}
Pitman J. \emph{Probability}. Springer Texts in Statistics, Springer: New York,
  Berlin, and Heidelberg, 1993.

\bibitem{Scott2011}
Scott LR. \emph{Numerical Analysis}. Princeton University Press, 2011.

\bibitem{PizDaint}
{CSCS Swiss National Supercomputing Center}. The {P}iz {D}aint supercomputer
  2015. \urlprefix\url{http://www.cscs.ch/computers/piz_daint/index.html}.

\bibitem{Sanan15}
Sanan P. {PGMRES} and {PIPECG} test data 2015.
  \urlprefix\url{https://bitbucket.org/psanan/pipe_daint_test}.

\bibitem{CsorgoFaraway1996}
Csorgo S, Faraway JJ. The exact and asymptotic distributions of {C}ram\'er-von
  {M}ises statistics. \emph{Journal of the Royal Statistical Society. Series B
  (Methodological)}  1996; :221--234.

\bibitem{RigdonBasu2000}
Rigdon SE, Basu AP. \emph{Statistical methods for the reliability of repairable
  systems}. Wiley: New York, 2000.

\bibitem{RuppWeinbubJungelGrasser15}
Rupp K, Weinbub J, J\"ungel A, Grasser T. Pipelined iterative solvers for
  graphics processing units. \emph{SIAM Journal on Scientific Computing}  2015;
  To appear.

\end{thebibliography}
\end{document}